
\font\lbf=cmbx10 scaled\magstep2

\def\bs{\bigskip}
\def\ms{\medskip}
\def\np{\vfill\eject}
\def\nl{\hfill\break}
\def\ni{\noindent}
\def\cl{\centerline}

\def\title#1{\cl{\lbf #1}\ms}
\def\stitle#1{\ms{\ni\bf #1}\par\nobreak\ms}

\def\Ref{\stitle{References}\ni}
\def\ref#1#2#3#4{#1\ {\it#2\ }{\bf#3\ }#4\nl}
\def\reff#1#2#3#4{#1\ {\it#2\ }{\bf#3\ }#4}
\def\ns{\kern-.33333em}
\def\CQG{Class.\ Qu.\ Grav.}
\def\GRG{Gen.\ Rel.\ Grav.}
\def\PL{Phys.\ Lett.}
\def\PR{Phys.\ Rev.}
\def\RPP{Rep.\ Prog.\ Phys.}

\def\address{\ms\cl{Max Planck Institut f\"ur Astrophysik}
		\cl{Karl Schwarzschild Stra\ss e 1}
		\cl{8046 Garching bei M\"unchen}
		\cl{Germany}\ms}
\def\me{\ms\cl{\bf Sean A.\ Hayward}\address}

\def\D{{\cal D}}
\def\G{{\cal G}}
\def\L{{\cal L}}
\def\R{{\cal R}}
\def\e{\varepsilon}
\def\t{\theta}
\def\s{\sigma}
\def\O{\Omega}
\def\half{{\textstyle{1\over2}}}
\def\third{{\textstyle{1\over3}}}
\def\twothird{{\textstyle{2\over3}}}
\def\quart{{\textstyle{1\over4}}}
\def\sixth{{\textstyle{1\over6}}}

\magnification=\magstep1

\def\o{\vert_S}
\def\sv{\varsigma}
\def\Sh{{\hat S}}
\def\gh{{\hat g}}
\def\th{{\hat\t}}
\def\sh{{\hat\s}}

\def\Eh{{\hat E}}
\def\Bh{{\hat B}}
\def\oh{\vert_\Sh}

\cl{\lbf On cosmological isotropy, quantum cosmology}
\cl{\lbf and the Weyl curvature hypothesis}
\me
\cl{24th June 1992}
\ms
{\bf Abstract.}
The increasing entropy, large-scale isotropy and approximate flatness
of the universe are considered in the context of signature change,
which is a classical model of quantum tunnelling in quantum cosmology.
The signature change hypothesis implies an initial inflationary epoch,
the magnetic half of the Weyl curvature hypothesis,
and a close analogue of the conformal singularity hypothesis.
Adding the electric half of the Weyl curvature hypothesis
yields, for a perfect fluid,
only homogeneous and isotropic cosmologies.
In the cosmological-constant case,
the unique solution is the Vilenkin tunnelling solution,
which gives a de Sitter cosmology.
\bs
Explaining the large-scale structure of the universe
is the ultimate goal of cosmology.
Although the dynamical evolution of the universe
is usually accepted as being determined by general relativity,
various particular properties of the universe can only be explained
in terms of particular initial conditions.
The most striking cosmological observations of this type are
the thermodynamic arrow of time,
the high isotropy of the universe at large scales,
and the approximate flatness of the universe.
Inflation [1--3] is widely regarded as providing a plausible explanation of
the last two observations,
albeit at a phenomenological rather than fundamental level.
\ms
It seems likely that only a quantum theory of gravity can adequately explain
the particular initial conditions of the universe.
Nevertheless, it is possible to make conjectures in classical terms
by appealing to the expected properties of a correct quantum theory of gravity.
In particular, there is the Weyl curvature hypothesis of Penrose [4--6],
which asserts that the Weyl tensor vanishes at the initial singularity,
and is intended to express a low initial entropy.
The initial cosmological singularity is predicted by general relativity
provided that positive-energy conditions are satisfied,
but creates various philosophical and mathematical difficulties,
which may be resolved either by abolishing the singularity
or by accomodating it in some way.
A suggestion in the latter direction,
due to Goode \& Wainwright [7--10] and in a simpler form to Tod [11--13],
is that the singularity be conformal or isotropic.
On this assumption, together with the Weyl curvature hypothesis,
R.\ Newman has shown that for a radiative perfect fluid,
the resulting universe must be homogeneous and isotropic [14].
This remarkable result raises the question of whether the assumptions
can indeed be justified by quantum gravity.
\ms
A promising candidate for a correct quantum theory of gravity is
quantum cosmology according to the Hartle-Hawking programme [15--18].
Unfortunately, this involves various mathematical and interpretational
problems,
and obtaining reliable predictions is rather difficult.
One way to obtain simple predictions is to take a classical limit of the
theory.
It might seem that this would merely yield classical cosmology according to
general relativity, but an alternative suggestion is that
the natural classical model of the
Hartle-Hawking proposal is a universe
whose signature is initially Riemannian but subsequently becomes Lorentzian.
In quantum cosmology, these are known as tunnelling solutions [19--22].
The change of signature can only occur at the Planck epoch,
since it is classically unstable [23].
\ms
Under the signature change hypothesis,
the initial singularity is abolished in favour of a compact non-singular
Riemannian region.
To make sense of the Weyl curvature hypothesis
in this context, it is natural to consider a vanishing Weyl tensor
at the signature-change junction,
since this will be observationally indistinguishable from
the original condition.
Then the following remarkable result is found:
the junction conditions for signature change imply
the magnetic half of the Weyl curvature hypothesis
and an analogue of the conformal singularity hypothesis.
Moreover, an initially inflationary universe is also predicted.
\ms
The derivation uses
the standard `3+1' decomposition [24--25] with unit lapse and zero shift,
where the metric $g_{ab}$ is decomposed into a spatial 3-metric $h_{ab}$
and a normal vector $n^a$ by $g_{ab}=h_{ab}-n_an_b$, $n_cn^c=-1$.
The second fundamental form may be divided into an expansion
$\t=\half h^{cd}\L_nh_{cd}$ and a traceless shear
$\s_{ab}=\half(\L_nh_{ab}-\third h_{ab}h^{cd}\L_nh_{cd})$,
where $\L_n$ is the Lie derivative along $n^a$.
The energy tensor $T_{ab}$ of the matter may be divided into an energy density
$\rho=T_{cd}n^cn^d$, a momentum $j_a=h_a^cT_{cd}n^d$,
a pressure $p=\third h^{cd}T_{cd}$ and a traceless stress
$\sv_{ab}=h_a^ch_b^dT_{cd}-\third h_{ab}h^{cd}T_{cd}$.
The Einstein equations are then
$$\eqalignno{
&0=3\s_{cd}\s^{cd}-2\t^2-3\R+6\rho,&(1)\cr
&0=3\D^c\s_{ac}-2\D_a\t-3j_a,&(2)\cr
&\L_nh_{ab}=2(\s_{ab}+\third\t h_{ab}),&(3)\cr
&\L_n\t=-\quart(2\t^2+3\s_{cd}\s^{cd}+\R+6p),&(4)\cr
&\L_n\s_{ab}=2\s_a^c\s_{bc}-\third\t\s_{ab}-\G_{ab}-\sixth\R h_{ab}+\sv_{ab},
&(5)\cr}
$$
where $\R$ is the Ricci scalar, $\G_{ab}$ the Einstein tensor and $\D_a$
the covariant derivative of $h_{ab}$.
The electric and magnetic parts of the Weyl tensor are
$$\eqalignno{
&E_{ab}=\half(\G_{ab}+\sixth\R h_{ab})
-\half(\L_n\s_{ab}-\third h_{ab}h^{cd}\L_n\s_{cd})
+\sixth\t\s_{ab},&(6)\cr
&B_{ab}=\e^{cd}{}_a(2\D_c\s_{bd}-h_{bc}\D^e\s_{de}+\twothird h_{bc}\D_d\t),
&(7)\cr}
$$
respectively, where $\e_{abc}$ is the alternating form of $h_{ab}$.
When considering signature change,
the Einstein equations in the Riemannian region
take a similar form with some sign changes,
and the Einstein equations at the junction itself
take the form of junction conditions [23].
\ms
An $n^a$-orthogonal 3-surface $S$ is isotropic,
both intrinsically and extrinsically, only if
$$
\s_{ab}\o=(\G_{ab}+\sixth\R h_{ab})\o=\sv_{ab}\o=0,\qquad
j_a\o=0,\eqno(8)
$$
since these quantities, if non-zero, yield preferred directions.
It follows from equations (1--8), using the Bianchi identities
$$
2\D^cE_{ac}=\D_a(p-\rho)-2j^c(\s_{ac}+\third\t h_{ac})-\D^c\sv_{ac},\qquad
\D^c\G_{ac}=0,
$$
that
$$
E_{ab}\o=B_{ab}\o=0,\qquad
\D_a\t\o=\D_a\R\o=\D_a\rho\o=\D_ap\o=0,
$$
so that the solution on $S$ is homogeneous if it is isotropic.
To deduce that the entire spacetime is homogeneous and isotropic given
the isotropy conditions (8) requires an assumption on the matter,
since a matter field with internal degrees of freedom
can break the symmetry of the gravitational field.
However, the standard Cauchy uniqueness theorem shows that
a homogeneous and isotropic cosmology arises from (8) for
a perfect fluid, given by
$$
T_{ab}=(\e+\pi)v_av_b+\pi g_{ab},\eqno(9)
$$
where $\e$ is the energy density, $\pi(\e)$ the pressure
and $v^a$ the flow direction:
$v_cv^c=-1$. No energy conditions are assumed here,
so that, for instance, a cosmological constant $\Lambda>0$ can be included
as the case $\e=-\pi=\Lambda$.
\ms
Before discussing the signature change model,
consider for comparison the conformal singularity model [7--14],
where a conformal transformation $g_{ab}=\O^2\gh_{ab}$ is assumed,
such that the initial singularity at $\O=0$ is a regular surface $\Sh$
in the conformal manifold.
The Weyl tensor is conformally invariant,
$\Eh_{ab}=E_{ab}$, $\Bh_{ab}=B_{ab}$.
Tod [12--13] shows that
the conformal second fundamental form vanishes at the singularity:
$$
\sh_{ab}\oh=0,\qquad\th\oh=0.\eqno(10)
$$
It follows that the magnetic half of the Weyl tensor
vanishes initially, $\Bh_{ab}\oh=0$ [7--14].
Adding the electric half $\Eh_{ab}\oh=0$ of the Weyl curvature hypothesis,
and assuming a radiative perfect fluid $\pi=\third\e$,
Newman [14] shows that only homogeneous and isotropic spacetimes result.
This is an intriguing result, though
the status of the conformal singularity hypothesis as a fundamental principle
is questionable, and like the Weyl curvature hypothesis
it presumably requires an appeal to the conjectured properties
of a correct quantum theory of gravity.
Happily, a very similar condition does arise in quantum cosmology,
or at least the signature-change model thereof, as follows.
\ms
The junction conditions for signature change are that the
second fundamental form vanishes at the junction $S$,
$$
\s_{ab}\o=0,\qquad\t\o=0,\eqno(11)
$$
with possible further conditions on the matter fields [20--23].
Such an initially stationary universe must have been strongly inflationary
during early epochs in order to expand to a size consistent with observation.
In particular, since
$$
\L_n\t=-\third\t^2-\s_{cd}\s^{cd}-\half(\rho+3p),
$$
initial inflation $\L_n\t\o>0$ occurs if and only if
$(\rho+3p)\o<0$. Conversely, if $\rho+3p>0$,
the universe would have rapidly collapsed to a caustic.
Thus there is a prediction of inflation and its
consequences for isotropy and flatness.
\ms
The junction conditions (11) are analogues of the conformal singularity
conditions (10), and various analogous results follow directly from
the Einstein equations (1--5) and Weyl identities (6--7),
without the technical difficulties caused by
the singular nature of the conformal field equations.
In particular, the magnetic half of the Weyl tensor vanishes initially,
$B_{ab}\o=0$, as does the momentum density, $j_a\o=0$.
For the particular case of a perfect fluid (9),
this means that the flow is orthogonal to $S$,
so that the traceless stress also vanishes,
$\sv_{ab}\o=0$.
Adding the electric half $E_{ab}\o=0$ of the Weyl curvature hypothesis
then yields the isotropy conditions (8),
and hence only homogeneous and isotropic cosmologies.
For instance, in the cosmological-constant case,
the Vilenkin tunnelling solution [19]
is the unique solution [23].
The Lorentzian part of this is de Sitter spacetime,
which is inflationary
and has a density parameter which approaches one asymptotically.
If this is taken as a model for the early evolution of the universe,
the observed density fluctuations could perhaps be explained by
initial quantum fluctuations, since the uncertainty principle forbids
a precisely zero Weyl tensor.
\ms
Given that the entropic motivation for the Weyl curvature hypothesis
appears to be satisfied by the magnetic half alone,
it is tempting to dispense with the electric half.
Also, the original formulation of the Weyl curvature hypothesis [4]
required the Weyl tensor to be initially only finite,
which is certainly the case in the context of signature change.
Thus I suggest simply using the signature change model as it stands,
since it can be justified as the classical limit of
Hartle-Hawking quantum cosmology.
Determining the possible initial conditions for the universe
is reduced to a problem in Riemannian geometry:
classify compact, simply-connected 4-manifolds with boundary,
with a non-singular Riemannian metric satisfying the Einstein equations
coupled to some well-defined matter source,
such that the boundary is totally geodesic [20--23].
(Compare with compact instantons [26--27].)
The Lorentzian spacetimes determined by the initial data on the junction
then constitute the set of predicted cosmologies,
and the spacetimes predicted by full quantum cosmology will presumably
be close to these.
If the Riemannian region is topologically a 4-hemisphere,
the only known solution is the Vilenkin solution,
and it may be conjectured to be unique, given the matter field.
This solution is at least known to be isolated [28].
\ms
In conclusion, the signature change hypothesis reveals various intriguing
connections between quantum cosmology, inflation, the conformal singularity
hypothesis and the Weyl curvature hypothesis,
and offers hope of a full explanation of the cosmological observations of
increasing entropy, large-scale isotropy and approximate flatness.
\np
\Ref
\ref{1. Linde A D 1984}\RPP{47}{925}
\ref{2. Blau S K \& Guth A H 1987 in}{300 Years of Gravitation}\ns
{ed Hawking \& Israel (Cambridge University Press)}
\ref{3. Linde A 1987}{ibid}{}{}
\ref{4. Penrose R 1979 in}
{General Relativity, an Einstein Centenary Survey}\ns{ed Hawking \& Israel
(Cambridge University Press)}
\ref{5. Penrose R 1981 in}{Quantum Gravity 2}\ns{ed Isham, Penrose \& Sciama
(Oxford University Press)}
\ref{6. Penrose R 1986 in}{Quantum Concepts in Space and Time}\ns
{ed Penrose \& Isham (Oxford University Press)}
\ref{7. Goode S W \& Wainwright J 1985}\CQG2{99}
\ref{8. Goode S W 1987}\GRG{19}{1075}
\ref{9. Goode S W 1991}\CQG8{L1}
\ref{10. Goode S W, Coley A A \& Wainwright J 1992}\CQG9{445}
\ref{11. Tod K P 1987}\CQG4{1457}
\ref{12. Tod K P 1990}\CQG7{L13}
\ref{13. Tod K P 1991}\CQG8{L77}
\ref{14. Newman R P A C 1991}{Twistor Newsletter}{33}{11}
\ref{15. Hartle J B \& Hawking S W 1983}\PR{D28}{2960}
\ref{16. Halliwell J J 1991 in}{Proc.\ Jerusalem Winter School
on Quantum Cosmology}\ns{ed Coleman, Hartle \& Piran (World Scientific)}
\ref{17. Hartle J B 1991}{ibid}{}{}
\ref{18. Page D N 1991 in}{Proc.\ Banff Summer Institute on
Gravitation}\ns{ed Mann \& Wesson (World Scientific)}
\ref{19. Vilenkin A 1982}\PL{117B}{25}
\ref{20. Halliwell J J \& Hartle J B 1990}\PR{D41}{1815}
\ref{21. Gibbons G W \& Hartle J B 1990}\PR{D42}{2458}
\ref{22. Fujiwara Y, Higuchi S, Hosoya A, Mishima T \& Siino M 1991}
\PR{D44}{1756}
\ref{23. Hayward S A 1992}\CQG9{1851}
\ref{24. Arnowitt R, Deser S \& Misner C W 1962 in}
{Gravitation: an Introduction to Current Research}\ns{ed Witten (Wiley)}
\ref{25. Fischer A E \& Marsden J E 1979 in}
{General Relativity, an Einstein Centenary Survey}\ns{ed Hawking \& Israel
(Cambridge University Press)}
\ref{26. Isham C J 1981 in}{Quantum Gravity 2}\ns{ed Isham, Penrose \& Sciama
(Oxford University Press)}
\ref{27. Pope C N 1981}{ibid}{}{}
\reff{28. Besse A L 1987}{Einstein Manifolds}\ns{(Springer-Verlag)}

\bye